\documentclass[12pt]{iopart}

\usepackage{graphicx}

\newcommand{\bef}{\begin{figure}}
\newcommand{\eef}{\end{figure}}

\newcommand{\be}{\begin{equation}}
\newcommand{\ee}{\end{equation}}
\newcommand{\bea}{\begin{eqnarray}}
\newcommand{\eea}{\end{eqnarray}}

\begin{document}


\title{Energy dependence of inclusive photon elliptic flow in heavy-ion collision models}
\author{Ranbir Singh$^1$, Md. Nasim$^2$, Bedangadas Mohanty$^2$ and Sanjeev Singh Sambyal$^1$}
\address{$^1$ Physics Department,University of Jammu, Jammu 180001, India and $^2$Variable Energy Cyclotron Centre, Kolkata 700064, India}

\begin{abstract}
We present a comparison of inclusive photon elliptic flow parameter ($v_{2}$)
measured at RHIC and SPS high energy heavy-ion collision experiments to
calculations done using the AMPT and UrQMD models. The new results 
discussed includes the comparison of the model calculations of photon
$v_{2}$ to corresponding measurements at the forward rapidities. 
We observe that the AMPT model which includes partonic interactions and quark coalescence 
as a mechanism of hadronization is in good agreement with the
measurements even at forward rapidities (2.3 $<$ $\eta$ $<$ 3.9) at
RHIC as was previously observed for measurements at midrapidity. 
At the top SPS energy the contribution from partonic effects are
smaller than that at RHIC energy, based on the comparison of the
measured  photon $v_{2}$ with those from the AMPT default and UrQMD model calculations. 
We find that if the measurements in RHIC beam energy scan (BES) and LHC energies
would require an energy dependent partonic cross section in the AMPT models, then the observed 
longitudinal scaling of $v_{2}$ at top RHIC energies (19.6--200 GeV) will be violated.
We also discuss the relation  between the inclusive photon $v_{2}$ and those of their parent 
$\pi^{0}$'s for the beam energies of 7.7 GeV to 2.76 TeV. The model results show that the
transverse momentum ($p_{\mathrm T}$) integrated $v_{2}$ of $\pi^{0}$ is larger by about 44\% relative 
to those of the inclusive photons. Finally we present the expectations of inclusive photon $v_{2}$ for 
the RHIC beam energy scan (BES) program and LHC from the transport models, so that they can be 
compared to corresponding measurements using the data already collected at RHIC and LHC.  
%
\end{abstract}
\maketitle
\section{Introduction}
\label{intro}
Elliptic flow ($v_{2}$) measurements are believed to provide information on the
dynamics of the system formed in the heavy-ion collisions~\cite{rhicflow,flow1}.
Within the framework
of a hydrodynamical approach, $v_{2}$ is found to be sensitive to the equation
of state of the system formed in the collisions~\cite{flow2}.
Measurements of elliptic flow at the forward rapidity have revealed an interesting
observation of longitudinal scaling of $v_{2}$ when plotted as a function of
pseudorapidity ($\eta$) shifted by the beam
rapidity ($y_{\mathrm {beam}}$)~\cite{phobosflow,nasim1}.
With the upcoming  new measurements in the beam
energy scan (BES) program at RHIC~\cite{starbes} and
higher energies at LHC, this scaling can be put
to a further test. At rapidities where these measurements are essential for studying
such scaling at RHIC and LHC we have mostly multiplicity detectors in both the colliders.
For example, in STAR experiment and ALICE there exists a photon multiplicity detector (PMD) in the range
2.3 $<$ $\eta$ $<$ 3.9~\cite{pmdstar,pmdalice}. In fact the azimuthal anisotropy
measurements using photons measured in PMD for S+Au collisions at 200 AGeV were the first
observation of collectivity at SPS energies~\cite{wa93}.
In this paper we concentrate on the $v_{2}$ of inclusive photons measured
by such a detector and discuss the limitations and possibilities it could offer
to understand the longitudinal scaling of $v_{2}$ in BES and LHC energies.

The $v_{2}$ is the 2nd Fourier coefficient of the particle azimuthal angle
($\phi$) distribution with respect to the reaction plane angle ($\Psi$).
Where the $\Psi$ is the angle subtended by the plane containing the impact
parameter vector and the beam direction (usually considered as the Z-axis)
with the X-axis~\cite{art}. Mathematically we can write,
\begin{equation} \frac{dN}{d\phi} \propto
1+2 v_2\cos(2(\phi - \Psi)). \end{equation}
For a given rapidity/centrality window the second coefficient or the
elliptic flow parameter is
\begin{equation}
v_{2}=\langle\cos(2(\phi-\Psi))\rangle.
\end{equation}
We present a study of the $v_{2}$ for inclusive photon using above approach
of measurement from transport models in heavy-ion collisions like Ultra relativistic
Quantum Molecular Dynamics (UrQMD)~\cite{urqmd} and A Multi Phase
Transport (AMPT) model~\cite{ampt}. Since more than
90\% of the inclusive photons come from the decay of $\pi^{0}$ mesons, we will first
study how the decay effect decreases the measured $v_{2}$ for photons relative to
those of the parent $\pi^{0}$. PMD like detectors have limitations in the form of
finite particle counting efficiency and also the purity of the photon sample~\cite{pmdmult}.
We will discuss the effect of these on the measured $v_{2}$. There already exists
measurements of $v_{2}$ of inclusive photons at SPS by the WA98 experiment~\cite{wa98flow}
and at RHIC by the STAR experiment~\cite{rashmi2008} and PHENIX experiment~\cite{phenix}.
Using two different approaches in the AMPT model: AMPT default and AMPT string
melting (AMPT-SM), we will discuss the relevance of partonic interactions and
partonic coalescence as a mechanism of hadronization to inclusive photon $v_{2}$
measurements in the above experiments. The models are also used to predict the
magnitude of $v_{2}$ expected at RHIC BES and LHC energies.

Before we proceed further, we discuss very briefly the main features of the two models,
the results of which are discussed in this paper. The UrQMD model is based on a
microscopic transport theory where the phase space description of the reactions
play an important role. The hadrons propagate on a classical trajectory
undergoing stochastic binary scattering, color string formation and resonance decay.
The model incorporates baryon-baryon, meson-baryon and meson-meson interactions.
The $v_{2}$  obtained from the UrQMD model will provide information on the
contribution from the hadronic interactions.
The AMPT model uses the same initial conditions as in
Heavy Ion Jet Ion Interaction Generator (HIJING)~\cite{hijing}.
Then the mini-jet partons are made to undergo scattering before they are allowed to fragment
into hadrons. The string melting (SM) version of the AMPT model is based on the idea that
for energy densities beyond a critical value of  about 1 GeV/$fm^{3}$, it is difficult to
visualize the coexistence of strings (or hadrons) and partons. Hence the need to melt the
strings to partons. This is done by converting the mesons to a quark and anti-quark pair,
baryons to three quarks (through intermediate diquark process) etc. The scattering of
the quarks are based on parton cascade ZPC~\cite{ampt}. Once the interactions stop,
the partons then hadronizes through the mechanism of parton coalescence.
The interactions between the mini-jet partons in AMPT model and those between partons in
the AMPT-SM model could give rise to a substantial $v_{2}$. The results from AMPT-SM would
indicate the contribution of partonic interactions to the observed $v_{2}$. Recently
it has been shown that while a parton-parton interaction cross section of 10 mb is
needed to explain the observed $v_{2}$ at midrapidity for top RHIC energies,
a much smaller cross section of 1.5 mb is required to explain the observed $v_{2}$
at midrapidity for the LHC energies~\cite{xuko}.
Hence it will be interesting to see what happens to the longitudinal scaling of $v_{2}$
if the requirement of parton-parton cross section changes as a function of center of mass
energy ($\sqrt{s_{NN}}$). In this paper we have generated events using the UrQMD (ver.2.3 ) 
and AMPT (ver. 1.25t3) (with HIJING (ver. 1.35)) event generators at various beam energies 
results of which are presented in this paper. The default conditions are used, For RHIC energies
the parton-parton cross section in AMPT-SM is taken to be 10 mb and for the LHC energy
it is taken as 1.5 mb.

The paper is organized as follows. In the next section we show a comparison between
$v_{2}$ of inclusive photon and $\pi^{0}$ as a function of $p_{\mathrm T}$,
$\eta$ and collision centrality. The energy dependence of the difference in $v_{2}$ of
inclusive photon to those from $\pi^{0}$ are also discussed. Then we present a
brief discussion on the effect of finite efficiency and purity of photon counting
on inclusive photon $v_{2}$ measurements using a simple model based approach.
In section III we compare the inclusive photon measurements at SPS and RHIC energies
to results from UrQMD and AMPT models. In section IV we present the expectations
of inclusive photon $v_{2}$ at RHIC BES and LHC energies. In the end of the section we
provide a discussion on longitudinal scaling of inclusive photon $v_{2}$. Finally we
summarize our study in section V.

\section{photon versus neutral pion $v_{2}$ at RHIC and LHC}

About 90\% of the inclusive photon come from decay of $\pi^{0}$ mesons~\cite{pmdmult}.
In this section we discuss how the $v_{2}$ of photons and $\pi^{0}$
compares as a function of collision centrality, $p_{T}$ and $\eta$
at $\sqrt{s_{\mathrm {NN}}}$ = 2.76 TeV (chosen
as a representative case) for Pb+Pb collisions at forward rapidity (2.3 $<$ $\eta$ $<$ 3.9).
We discuss the results at forward rapidity as in this region the actual
measurements will be carried out in the experiments. We study the effect using AMPT
model with two versions - default and SM.
\bef
\begin{center}
\includegraphics[scale=0.4]{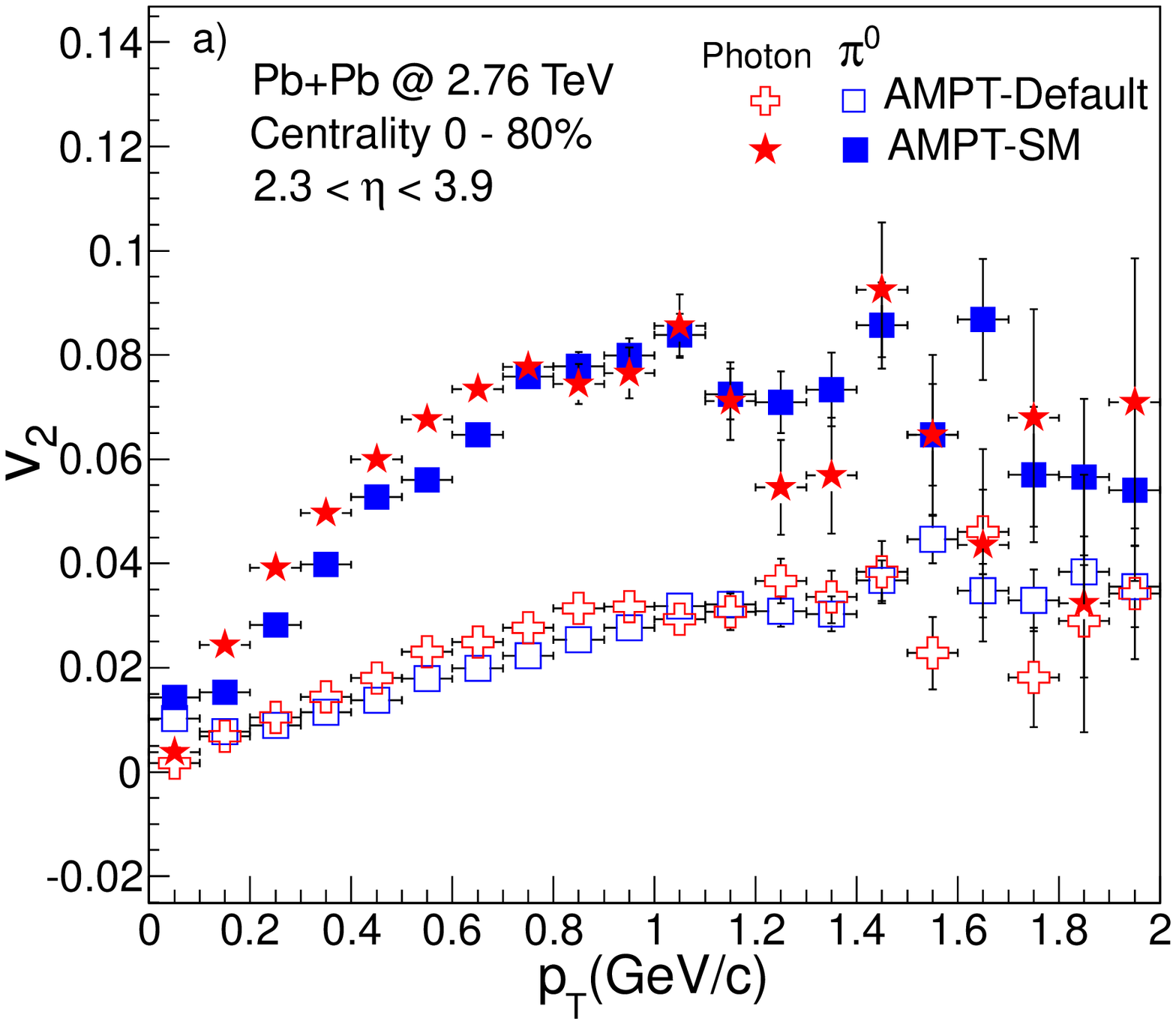}
\includegraphics[scale=0.4]{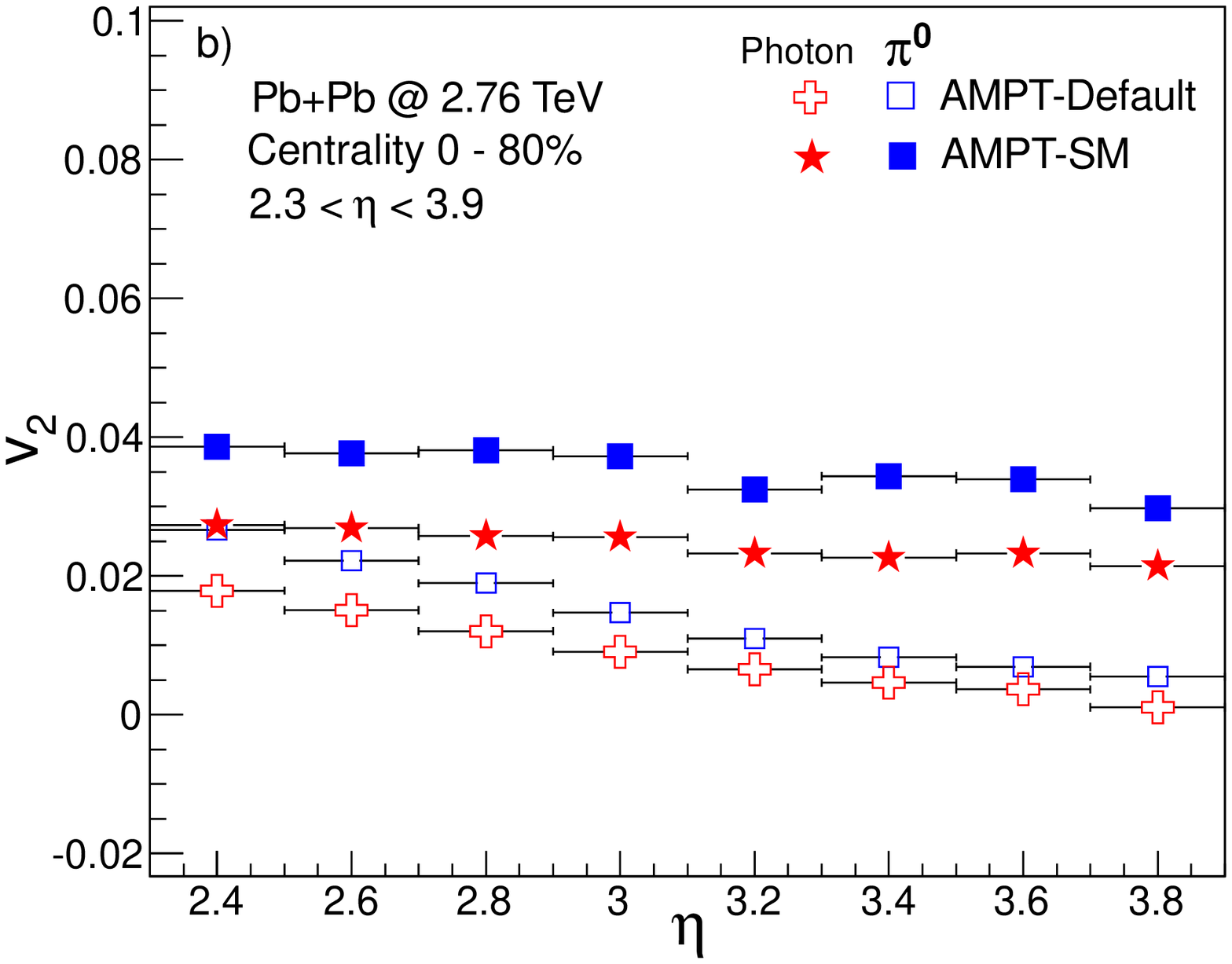}
\includegraphics[scale=0.4]{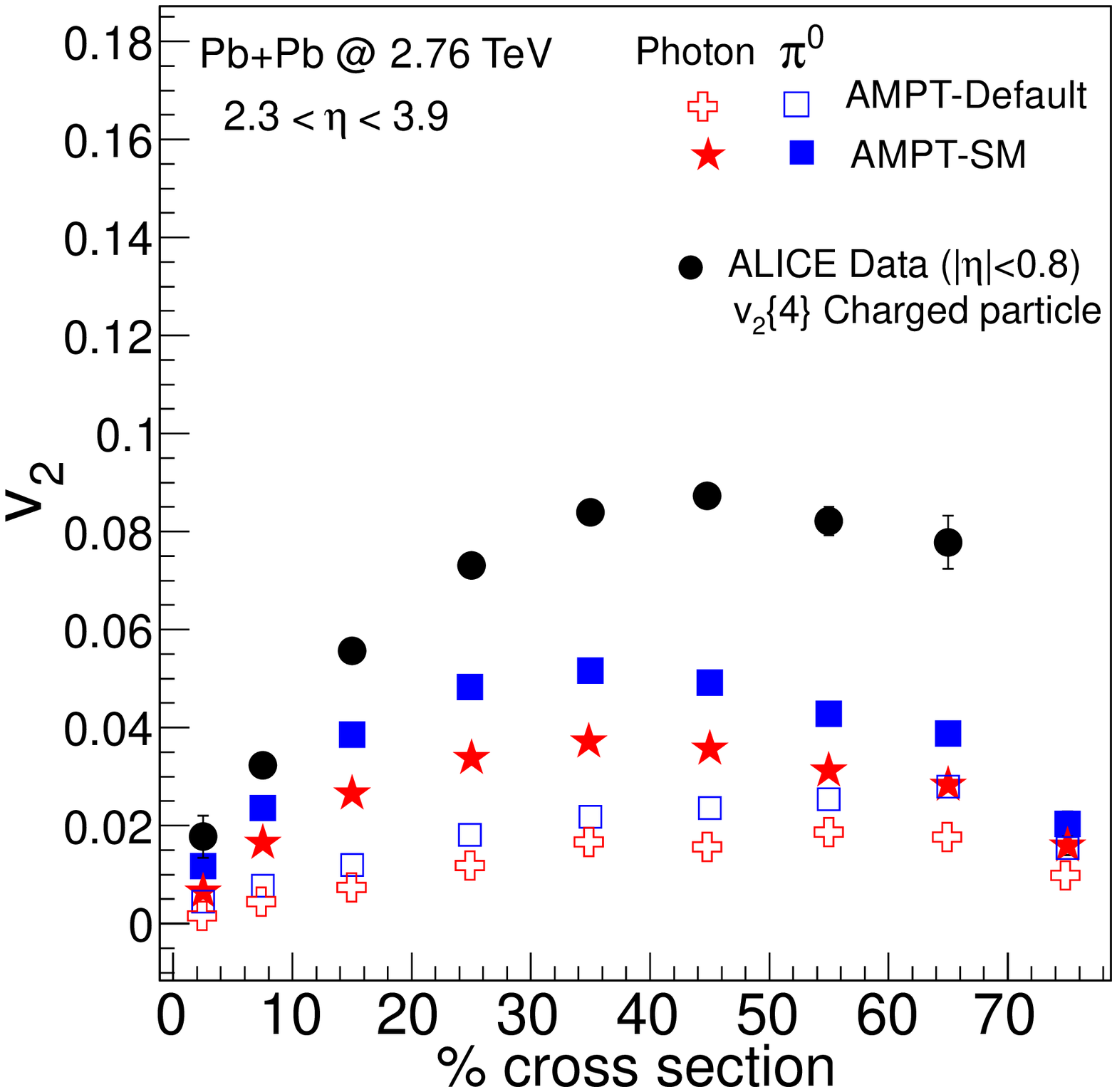}
\caption{(Color online) Elliptic flow ($v_{2}$) as a function $p_{\mathrm T}$ (a),
 $\eta$ (b) and \% cross section (c) for inclusive photon
and $\pi^{0}$ from AMPT-default and AMPT-SM in 0-80\% Pb+Pb collisions
for 2.3 $<$ $\eta$ $<$ 3.9 at $\sqrt{s_{\mathrm {NN}}}$ = 2.76 TeV. Also shown are the 
$v_{2}$ results obtained from 4-particle cumulant method for the inclusive charged hadrons 
from the ALICE experiment~\cite{alicechv2} at midrapidity for comparison. It may be noted
that the maximum value of $p_{\rm T}$ integrated $v_{2}$ at midrapidity for charged hadrons
for Pb+Pb collisions at $\sqrt{s_{\mathrm {NN}}}$ = 2.76 TeV is 0.087 $\pm$  0.002 (stat) 
$\pm$ 0.003 (syst)~\cite{alicechv2}.}
\label{pi0gamma}
\end{center}
\eef

Figure~\ref{pi0gamma} shows the comparison of $v_{2}$ for photons and $\pi^{0}$.
The Fig.~\ref{pi0gamma}a shows the results as a function of $p_{T}$. We see that for both
AMPT-SM and AMPT default the $v_{2}$ for photons and $\pi^{0}$ are comparable for
minimum bias Pb+Pb collisions. The magnitude of $v_{2}$ from AMPT-SM is
higher than that from AMPT-default. The Fig.~\ref{pi0gamma}b shows the $v_{2}$ as a
function of $\eta$. The $p_{T}$ integrated $v_{2}$ of photon is smaller than
that for $\pi^{0}$ for minimum bias Pb+Pb collisions. The Fig.~\ref{pi0gamma}c shows the
results as a function of collision centrality denoted by \% cross section.
The collision centrality could very well be denoted by the number of
participating nucleons ($N_{\mathrm {part}}$), calculated commonly
using Glauber Model.
The nucleon-nucleon cross sections are 42 and 65 mb for
$\sqrt{s_{\mathrm {NN}}}$ = 200 GeV and 2.76 TeV respectively.
At $\sqrt{s_{\mathrm {NN}}}$ = 2.76 TeV for Pb+Pb collisions, typically 0-5\%, 5-10\%, 10-20\%, 20-30\%, 30-40\%,
40-50\%, 50-60\%, 60-70\%, 70-80\%  cross section have the
corresponding average $N_{\mathrm {part}}$ values as
382$\pm$3, 330$\pm$4, 260$\pm$4, 186$\pm$4, 129$\pm$3, 85$\pm$3, 53$\pm$2, 30$\pm$1, 16$\pm$1 respectively.
While for Au+Au collisions at $\sqrt{s_{\mathrm {NN}}}$ = 200 GeV, the 0-5\%, 5-10\%, 10-20\%, 20-30\%, 30-40\%,
40-50\%, 50-60\%, 60-70\%, 70-80\% cross section have the
corresponding average $N_{\mathrm {part}}$ values  as
350$\pm$3, 301$\pm$7, 236$\pm$8, 168$\pm$10, 116$\pm$11, 76$\pm$10, 48$\pm$9, 28$\pm$7, 14$\pm$5 respectively.
As seen in the case for $\eta$ dependence, the $p_{T}$ integrated $v_{2}$ in the 2.3 $<$ $\eta$ $<$ 2.9 for photons are found to be
smaller than $\pi^{0}$. Both models display the expected trends of $v_{2}$ as a function
of centrality, with $v_{2}$ being smaller for central collisions. The lower value of $p_{T}$
integrated $v_{2}$ of photon compared to those for $\pi^{0}$
although $v_{2}$($p_{\mathrm {T}}$)
are comparable is due to photon $p_{\mathrm T}$ spectra having a smaller mean transverse
momentum compared to corresponding value for $\pi^{0}$. So for a multiplicity detector
as in STAR or ALICE experiments, without possibility of having $p_{T}$ information, the
measured photon $v_{2}$ would reflect a lower value than expected from the parent $\pi^{0}$.
Similar conclusions are observed for other energies (results from which are presented below)
we have studied and so not discussed in detail. 
Also shown in  Fig.~\ref{pi0gamma}c are the results of inclusive
charged hadron $v_{2}$ using 4-particle cumulant method
at midrapidity for  Pb+Pb collision at $\sqrt{s_{\mathrm {NN}}}$ = 2.76 TeV from the ALICE experiment~\cite{alicechv2}. 
The inclusive charged hadron $v_{2}$ values at midrapidity are more than a factor 2 higher than the values of 
$v_{2}$ from inclusive photon of $\pi^{0}$ calculated using models at
the forward rapidity. These results suggest that one could expect a distinct rapidity 
dependence of $v_{2}$  at LHC energies, with the $v_{2}$ values
starting to decrease before or around $\eta$ = 2.3 units. 

Figure~\ref{diff} shows the difference in $v_{2}$ of $\pi^{0}$ compared to those from the
photons divided by the $v_{2}$ of photons as a function of beam energy from AMPT SM model
using minimum bias collisions within 2.3 $<$ $\eta$ $<$ 2.9. This relative fraction is
observed to be almost constant as a function of $\sqrt{s_{\mathrm {NN}}}$ and has a value of
around 0.44. Similar conclusions are observed from other transport models like
AMPT-default and UrQMD and hence not presented here.
\bef
\begin{center}
\includegraphics[scale=0.4]{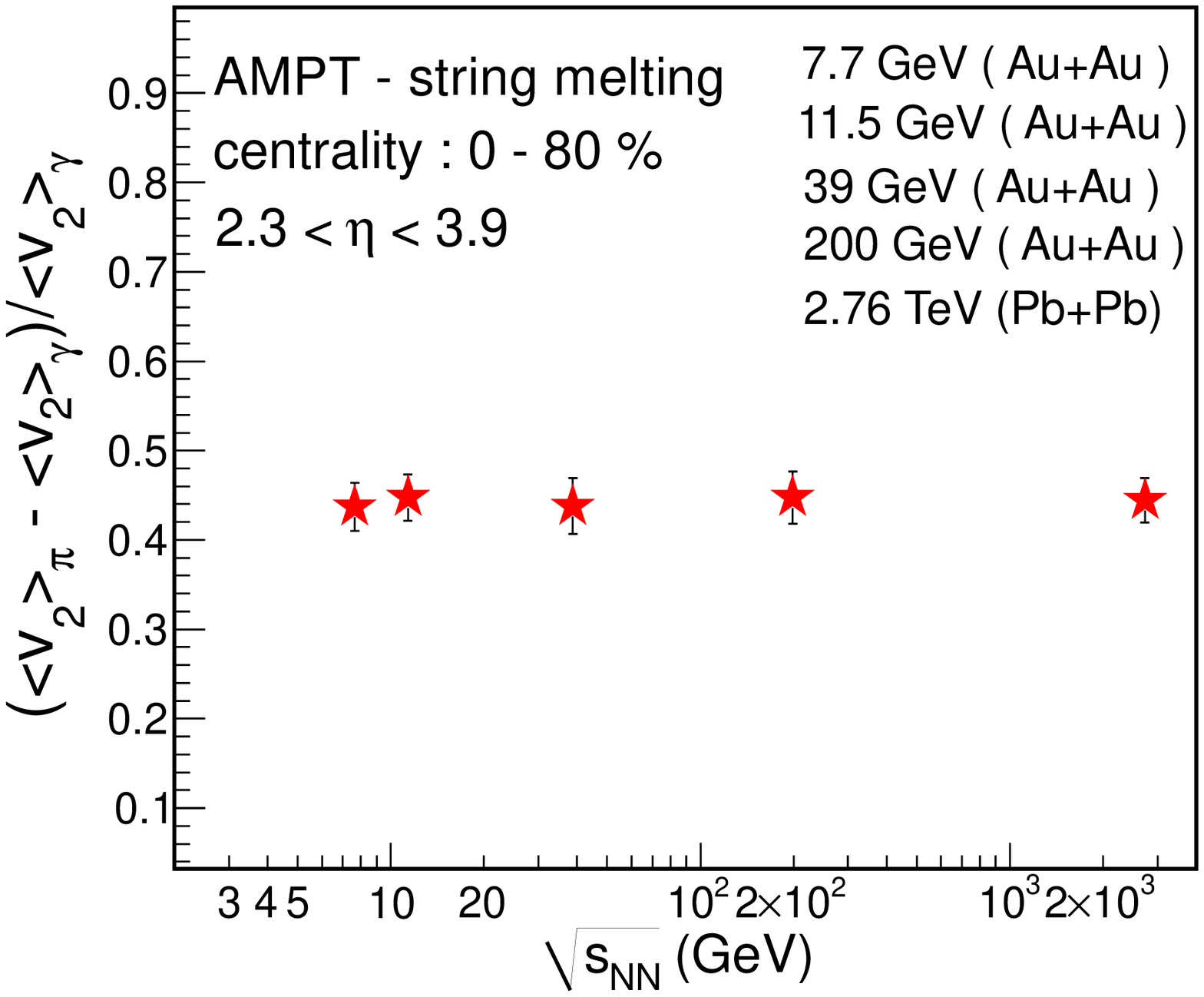}
\caption{(Color online) Ratio of the difference in average elliptic flow ($\langle v_{2} \rangle$)
for $\pi^{0}$ and inclusive photon to the $\langle v_{2} \rangle$ of photons as a function
of beam energy. The results as shown for 0-80\% nucleus-nucleus collisions for 2.3 $<$ $\eta$ $<$ 3.9 using AMPT-SM model.}
\label{diff}
\end{center}
\eef

The photon multiplicity detectors are affected by finite efficiency of photon
counting and purity of the detected photon sample. We have estimated the effect of
this on the measured $v_{2}$ of photons. For studying the effect of efficiency, we
have randomly removed photons from an event in the measured $\eta$ range by a factor
which depends on the photon counting efficiency. The photon counting efficiency
is defined as the ratio of the final number of photons detected to the initial number
of photons in an event.  This number is varied from 30\% to 95\% to see the
effect on $v_{2}$.
The results are shown in Fig.~\ref{effpur}.  We observe that the change in efficiency
does not affect the measured $v_{2}$. The typical value of efficiency for a realistic
PMD is about 60\%~\cite{pmdmult}. From $v_{2}$ measurement perspective this is a trivial
result as long as the efficiency does not have a azimuthal angle dependence.

The purity of the photon sample in multiplicity detectors are dominantly affected
by charged particles mimicking a photon signal. They are usually reduced by applying
certain thresholds on the signal deposited. Here the effect is studied by
adding a certain number of charged hadrons to the photon sample, yet maintaining
the same number of initial photons in an event. The number to be added is called as the
contamination and is related to the purity as (1-purity). Where purity is defined
as the ratio of number of photons in an events to total number of photons and
charged particles considered as photons in the event. The results are shown
in Fig.~\ref{effpur}. One observes that the purity affects the measured
$v_{2}$ of photons significantly. The measured $v_{2}$ increases relative to the photon
$v_{2}$ as the purity of the photon sample decreases.  Hence it is suggested that for
photon $v_{2}$ measurements in experiments, stability of the results needs to be checked
by varying the thresholds used to discriminate photon and hadrons falling on the detector.
Further in the experiments a detailed simulation using realistic detector configurations
can be used to understand the effects. Typical values of purity encountered is
around 60-70\%~\cite{pmdstar,pmdalice,wa98flow}, which as per the present study, could lead to a difference in actual and
measured $v_{2}$ of the order of 15\%.

\bef
\begin{center}
\includegraphics[scale=0.4]{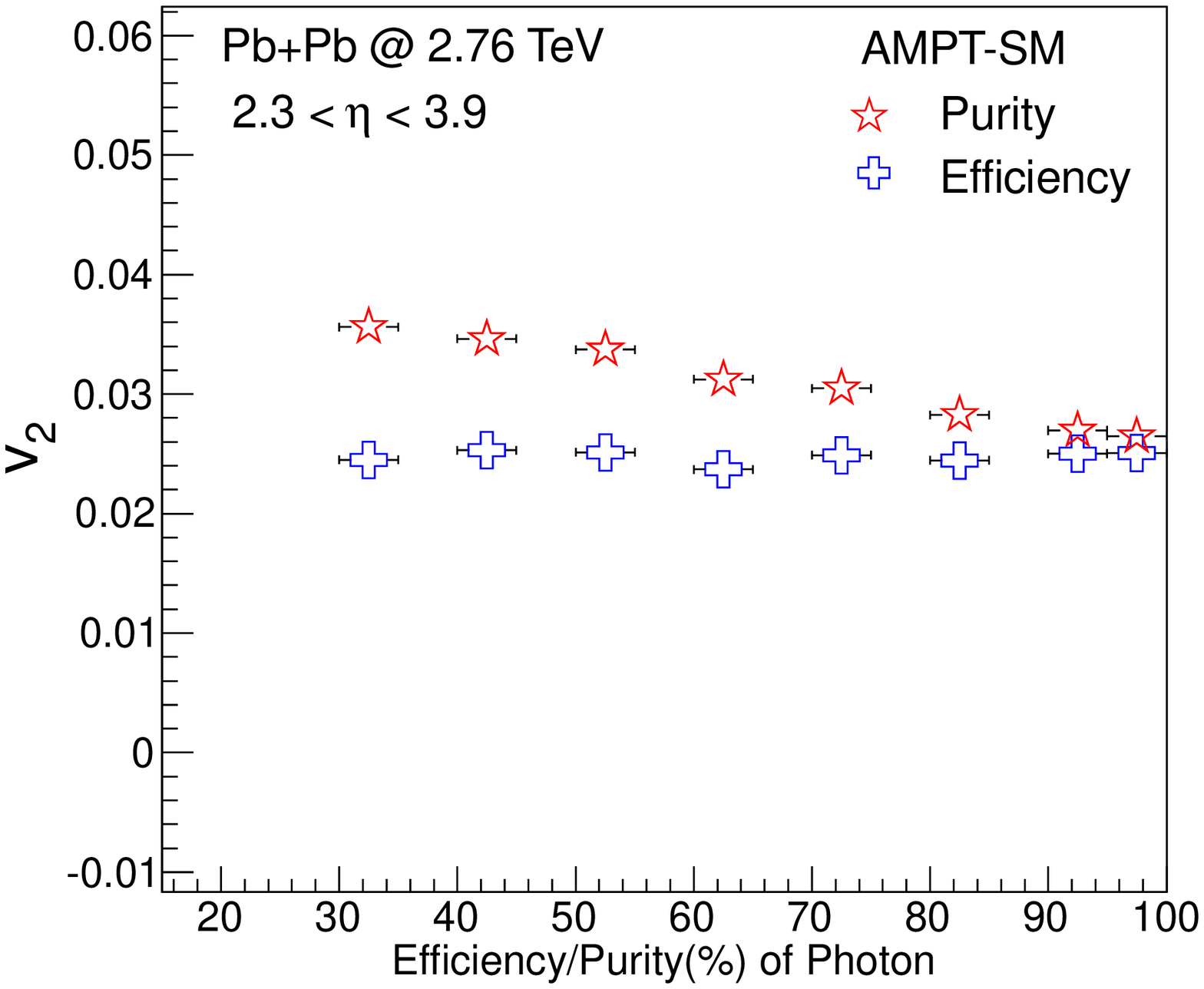}
\caption{(Color online) $v_{2}$ of inclusive photon as a function
of photon counting efficiency and purity (in \%) using AMPT-SM in 0-80\% Pb+Pb
collisions for 2.3 $<$ $\eta$ $<$ 3.9 at $\sqrt{s_{\mathrm {NN}}}$ = 2.76 TeV.}
\label{effpur}
\end{center}
\eef

\section{Comparison of photon $v_{2}$ measurements at SPS and RHIC to models}

In this section we compare the inclusive photon $v_{2}$ measurements to three
different transport model calculations. While AMPT default and UrQMD does not
include any partonic effects, AMPT-SM includes such a contribution. The difference
between AMPT default and UrQMD could lie in the treatment of initial and final state
re-scattering effects. A comparison of data with results from AMPT default and
UrQMD relative to those from AMPT-SM will help us to understand the contribution
to $v_{2}$ from partonic effects. The comparison of data with results from AMPT
default and UrQMD will help understand the role of re-scattering. Further we
discuss the effect of partonic cross section on $v_{2}$ measurements by choosing
their values to be 3, 6 and 10 mb in AMPT-SM model and comparing the results to
existing inclusive photon $v_{2}$ measurements at SPS and RHIC energies. With a
factor of 10 difference in beam energy, one would naively expect different
partonic cross section for SPS and RHIC energies which gives the best agreement with the
measured $v_{2}$.

\bef
\begin{center}
\includegraphics[scale=0.4]{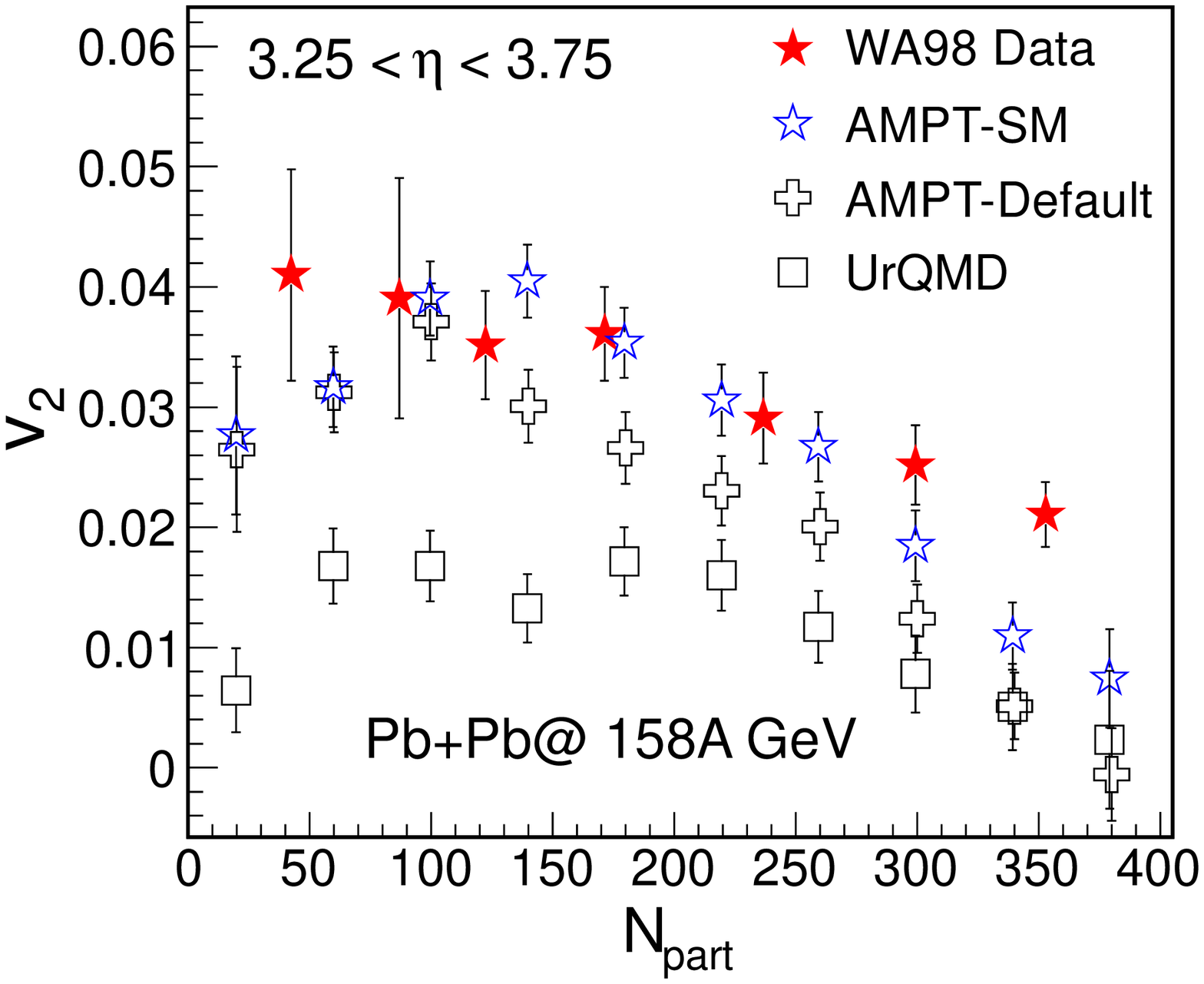}
\includegraphics[scale=0.4]{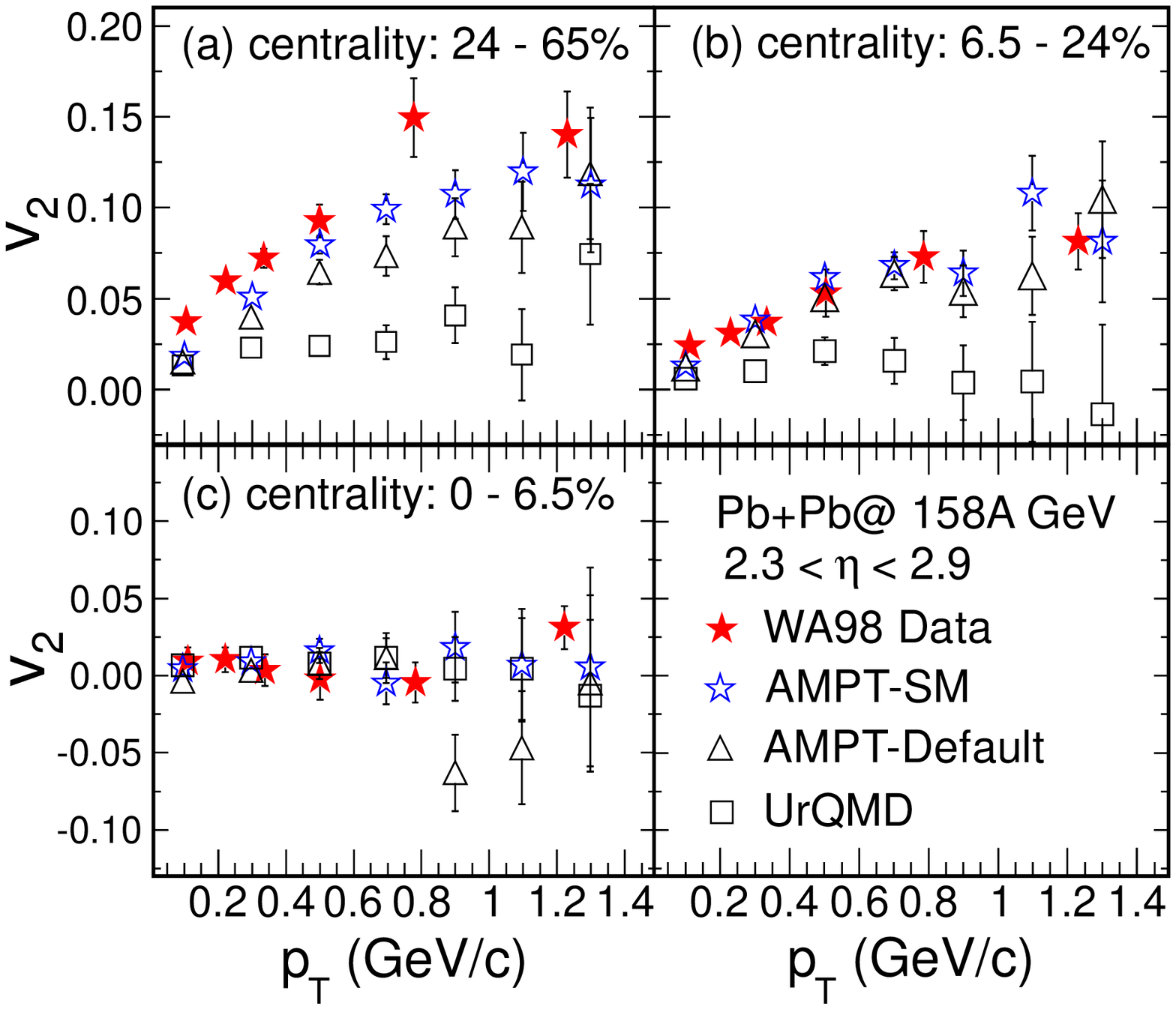}
\caption{(Color online) Upper panel: $v_{2}$ of inclusive photon as
a function of collision centrality for Pb+Pb collisions at laboratory
energy of 158 AGeV as measured by the photon multiplicity detector
in WA98 experiment~\cite{wa98flow}. The results are compared to corresponding values
from AMPT-SM, AMPT default and UrQMD models.
Lower panel: $v_{2}$ of inclusive photon as a function of $p_{T}$ for
three collision centrality as measured by the Pb-Glass Calorimeter in WA98
experiment~\cite{wa98flow}. The results are compared to the three transport
model calculations as mentioned above.}
\label{sps}
\end{center}
\eef

Figure~\ref{sps} shows the comparison of $v_{2}$ of inclusive photon as a function
of average number of participating nucleons and $p_{\mathrm T}$ for three different
centrality classes in Pb+Pb collisions at $E_{lab}$ = 158 AGeV measured by the
WA98 experiment ~\cite{wa98flow} at CERN SPS to various transport model results.
The centrality dependence results are obtained using a photon multiplicity detector
while the $p_{\mathrm T}$ dependence results were obtained from a Lead-glass calorimeter.
The error bars includes the systematic uncertainties, including the estimated effect of
charged hadron contamination to the measurements.
We observe that for this SPS top beam energy, in general,
$v_{2}$(UrQMD) $<$ $v_{2}$(AMPT-default) $<$ $v_{2}$(AMPT-SM). The results from AMPT-SM
having the closest agreement with the measurements both as a function of collision
centrality and $p_{T}$. However the  results from default version, which does not
include partonic effects are also not far off.

\bef
\begin{center}
\includegraphics[scale=0.4]{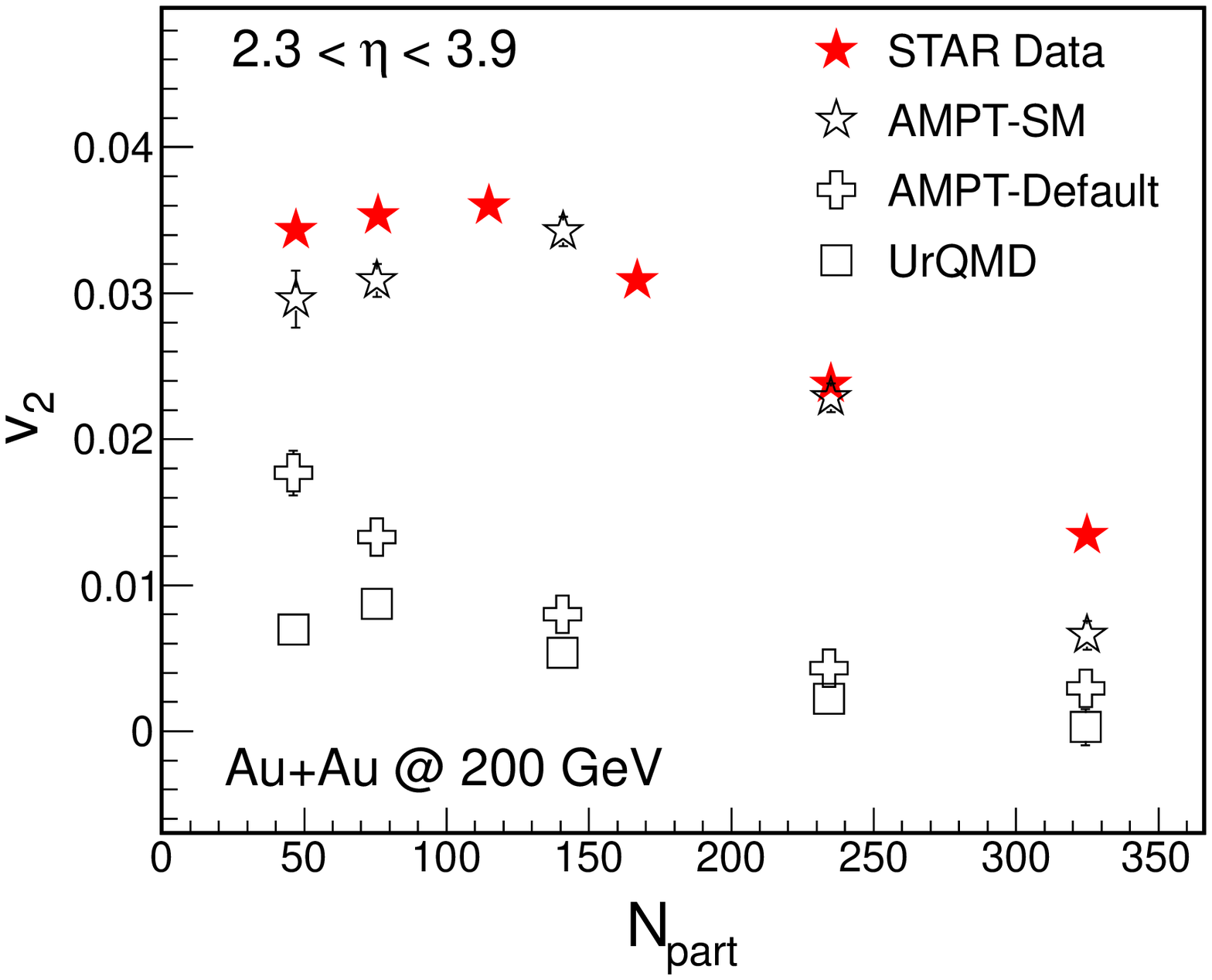}
\caption{(Color online) Preliminary $v_{2}$ of inclusive photon as
a function of collision centrality measured by the photon multiplicity
detector in STAR experiment at RHIC for Au+Au collisions at
$\sqrt{s_{\mathrm {NN}}}$ = 200 GeV~\cite{rashmi2008}.
The results are compared to transport model calculations
using AMPT-SM, AMPT default and UrQMD models.}
\label{rhic}
\end{center}
\eef

Figure~\ref{rhic} shows the comparison of preliminary result on $v_{2}$ of
inclusive photon as a function of collision centrality measured by the
STAR experiment at $\sqrt{s_{NN}}$ = 200 GeV using a photon multiplicity detector
~\cite{rashmi2008} to various transport model results.  While the
$v_{2}$ from AMPT default and UrQMD are comparable to each other, the $v_{2}$ from AMPT-SM
is much higher and comparable to the experimental measurements. This clearly shows that even
at forward rapidity in RHIC a substantial contribution of the measured $v_{2}$ could be
due to partonic interactions. Similar conclusions were derived from the midrapidity
measurements from inclusive charged particle $v_{2}$. However the model comparison to
top SPS energies results as shown in Fig.~\ref{sps} indicates that the partonic
contribution to $v_{2}$ is small relative to the case for RHIC. It may be noted that the
AMPT default and UrQMD results for RHIC energy as shown in Fig.~\ref{rhic} are comparable
while those for the SPS energy are different. This could be due to the different center of
mass rapidity range of the measurements. The SPS results are close to midrapidity in the
center of mass system while those for the RHIC are at forward rapidity.

\bef
\begin{center}
\includegraphics[scale=0.4]{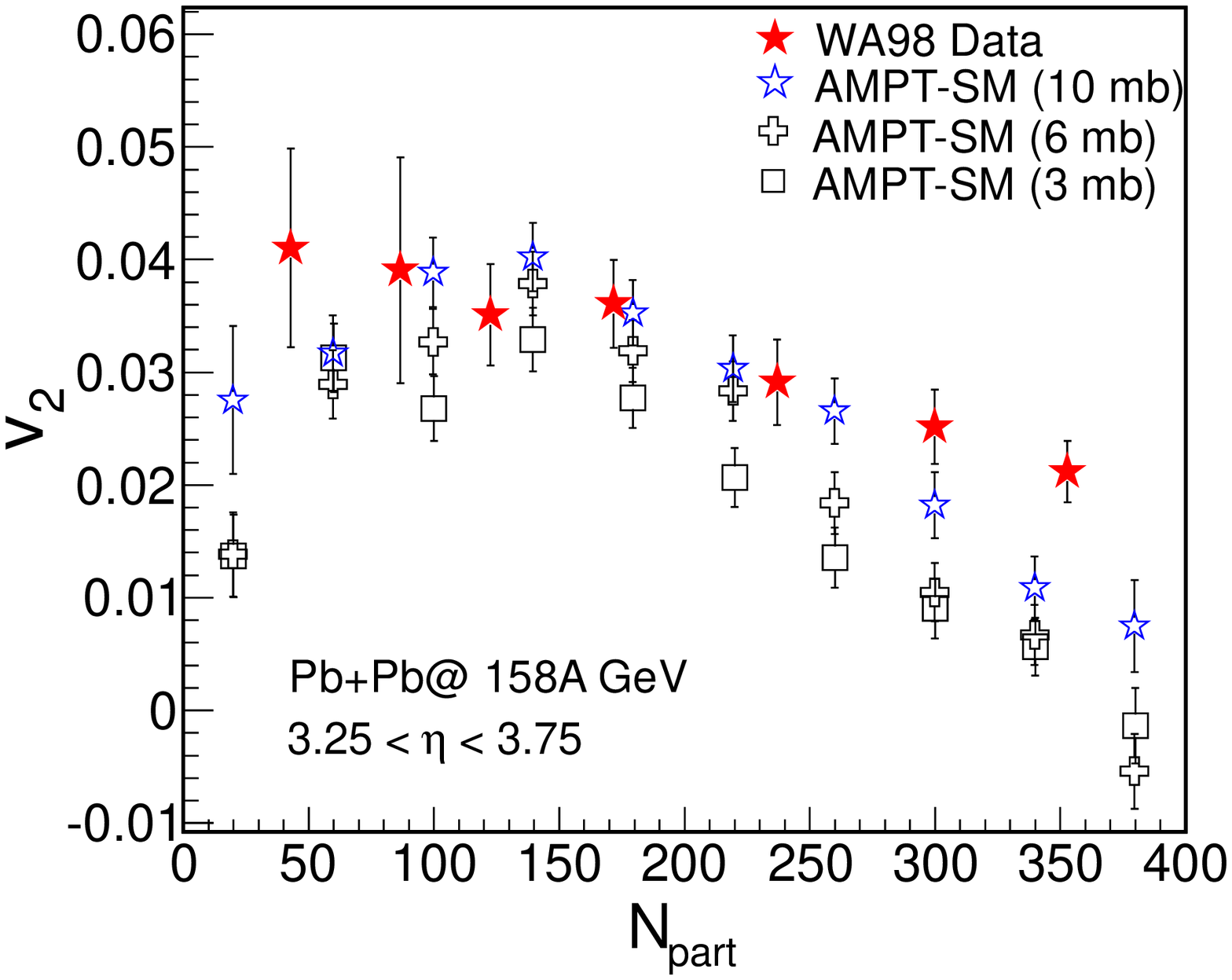}
\includegraphics[scale=0.4]{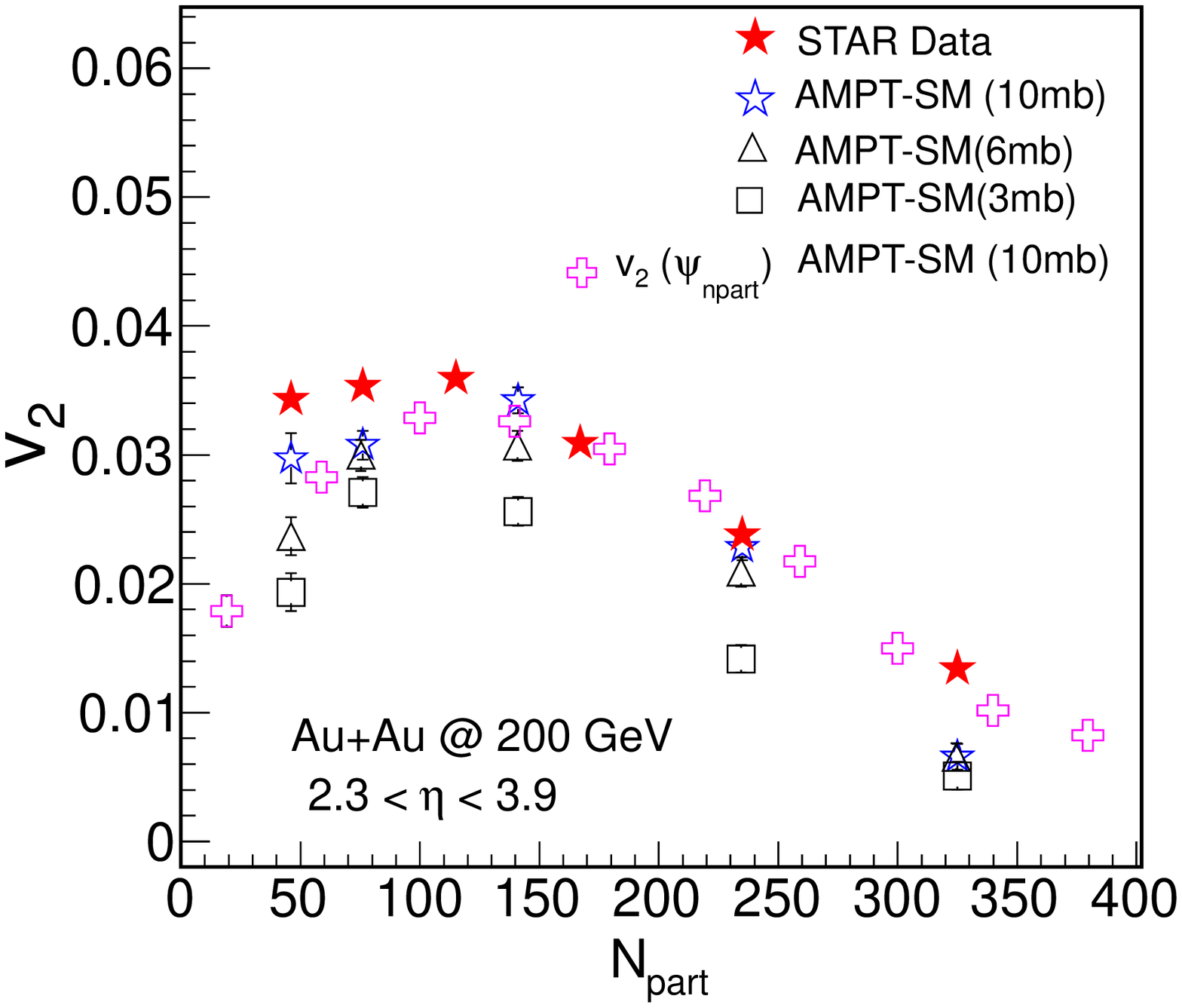}
\caption{(Color online) Upper panel: $v_{2}$ of inclusive photon as
a function of collision centrality for Pb+Pb collisions at laboratory
energy of 158 AGeV as measured by the photon multiplicity detector
in WA98 experiment~\cite{wa98flow}. The results are compared to corresponding values
from AMPT-SM with three different values of partonic cross sections.
Lower panel: Preliminary $v_{2}$ of inclusive photon as
a function of collision centrality measured by the photon multiplicity
detector in STAR experiment at RHIC for Au+Au collisions at
$\sqrt{s_{\mathrm {NN}}}$ = 200 GeV~\cite{rashmi2008}. 
The results are compared to AMPT-SM  model calculations
with three different values of partonic cross section. The value of $v_{2}$ with respect to 
participant plane $\Psi_{\rm npart}$ from AMPT-SM model is also shown to indicate the effect 
of flow fluctuations.}
\label{partoniccs}
\end{center}
\eef

Figure~\ref{partoniccs} shows the comparison of the centrality dependence of
inclusive photon $v_{2}$ measured at SPS and RHIC to the results from the AMPT-SM model with
three different partonic cross sections of 3, 6 and 10 mb. We find that
the $v_{2}$ values from the AMPT-SM model calculations increases with
increase in the value of the partonic cross section used for both at SPS
and RHIC energies. The data at both the energies seems to favor a large
partonic cross section of 10 mb. Due to the possibility of change in the
systems composition and the access to different phases (QGP or hadronic)
with change in beam energy, one would have expected the data at RHIC and
SPS to favor a different partonic cross section. However it must be noted
that while the results at SPS are close to the midrapidity range in the
center of mass frame, those at RHIC are at forward rapidity. It has been
that several bulk properties of matter at forward rapidity at RHIC is similar to
those at midrapidity at SPS~\cite{forwardmid}. New data from the RHIC BES
program at lower energies and higher LHC energies or in wide rapidity range will help
in studying the sensitivity of the measurements to change in partonic cross section
used in transport model calculations. 

Also shown in the Fig~\ref{partoniccs} for RHIC energies are the AMPT-SM calculation of $v_{2}$
using the participant plane ($\Psi_{\rm npart}$). The other model calculations uses the standard 
reaction plane ($\Psi$). One observes that the $v_{2}$ for central collisions obtained using participant plane 
from the model calculation are larger than the corresponding values obtained using the 
reaction plane. The agreement with the measurements are much better for the central 
collisions for $v_{2}$ obtained using $\Psi_{\rm npart}$. The $v_{2}$ calculated using reaction plane neglected the contributions from flow fluctuations~\cite{flowfluc}. 
It may also be noted that AMPT model calculations failed to explain the identified hadron $v_{2}$ at RHIC energies~\cite{ampt}.

\bef
\begin{center}
\includegraphics[scale=0.4]{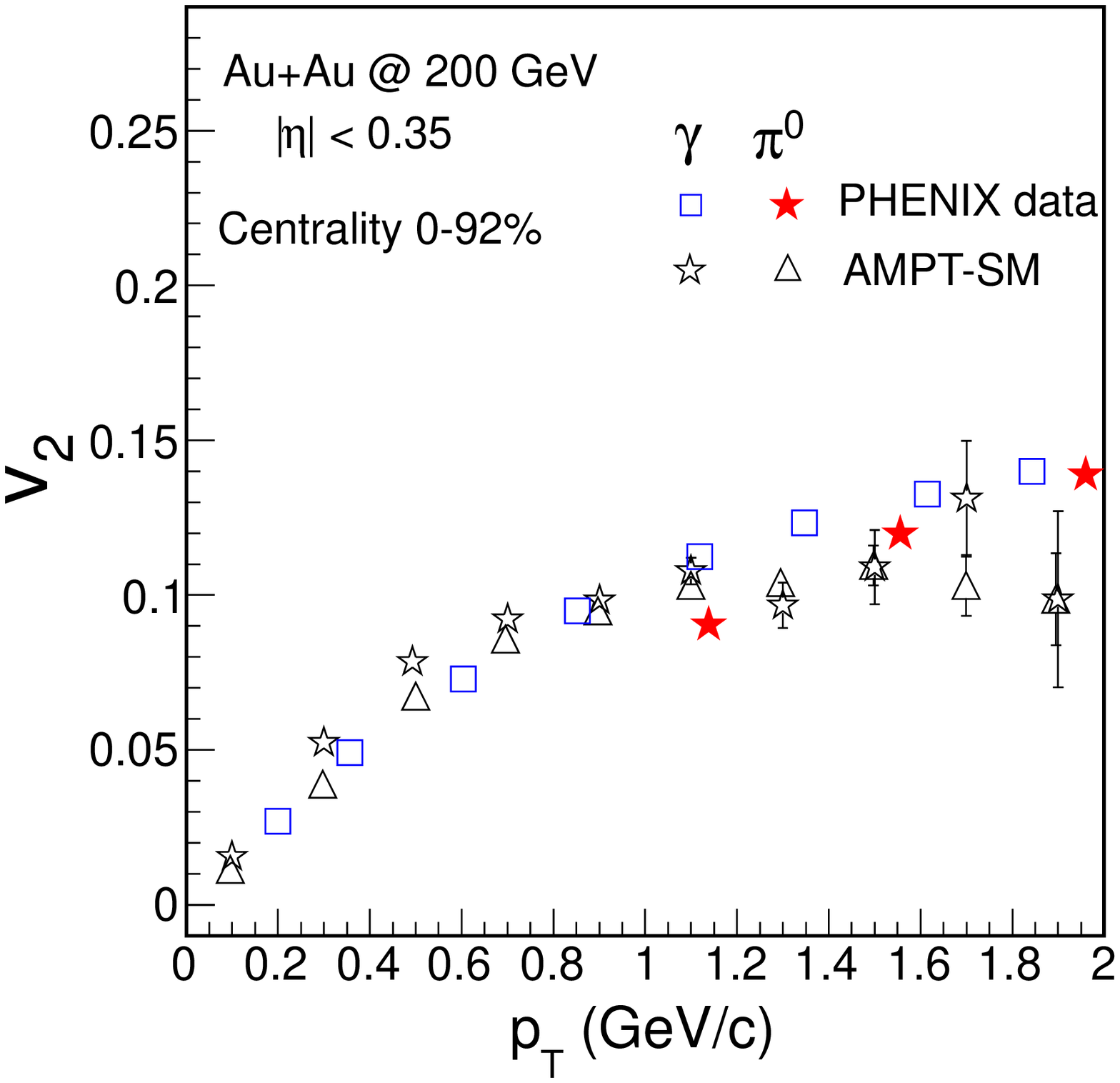}
\caption{(Color online) Inclusive photon and $\pi^{0}$  $v_{2}$ as a function
$p_{\rm T}$  at midrapidity for minimum bias Au+Au collisions at
$\sqrt{s_{\mathrm {NN}}}$ = 200 GeV~\cite{phenix}. The results are compared to
AMPT-SM model calculations. }
\label{phenix_ph}
\end{center}
\eef
Figure~\ref{phenix_ph} shows the inclusive photon and $\pi^{0}$ $v_{2}$ as a function of $p_{T}$ 
measured by PHENIX collaboration for Au+Au collisions at $\sqrt{s_{\mathrm {NN}}}$ = 200 GeV at 
midrapidity ($\mid \eta \mid <$ 0.35) for minimum bias collisions. We have compared the data 
to calculations from AMPT-SM model. We observe good agreement between the data and model results 
with calculations done using parton cross section of 10 mb. Both the mid rapidity and forward rapidity
RHIC results suggest that the data favors a large partonic cross section of 10 mb.

\section{Expectation of photon $v_{2}$ at RHIC Beam Energy Scan Program and LHC}

RHIC has recently complete a Beam Energy Scan (BES) program, by collecting
data for the $\sqrt{s_{\mathrm {NN}}}$ = 7.7 - 200 GeV, in the year 2010-2011~\cite{starbes}.
The main goal of the program is to study the structure of the QCD phase diagram~\cite{bm}.
Photon $v_{2}$ measurements at forward rapidities in this program together with
those at forward rapidity will provide information on the longitudinal scaling
of $v_{2}$. It may be mentioned that photon multiplicity measurements
at RHIC were used to study for the first time identified particle longitudinal
scaling behavior of the multiplicity distributions~\cite{pmdmult}. The longitudinal
scaling for inclusive charged particles were earlier reported by PHOBOS~\cite{phobos_sc} and BRAHMS~\cite{brahms_sc} 
experiments at RHIC.

\bef
\begin{center}
\includegraphics[scale=0.4]{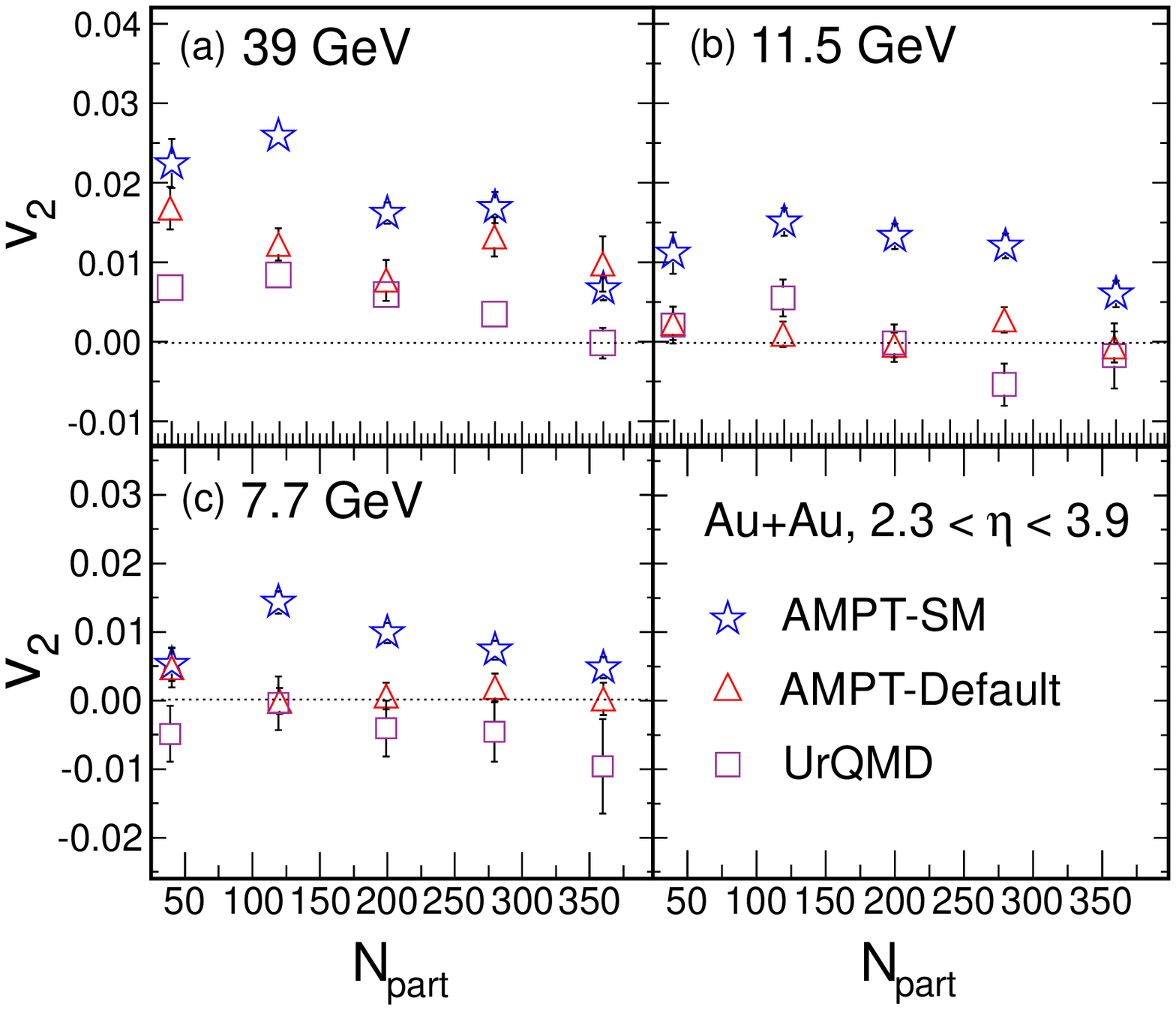}
\includegraphics[scale=0.4]{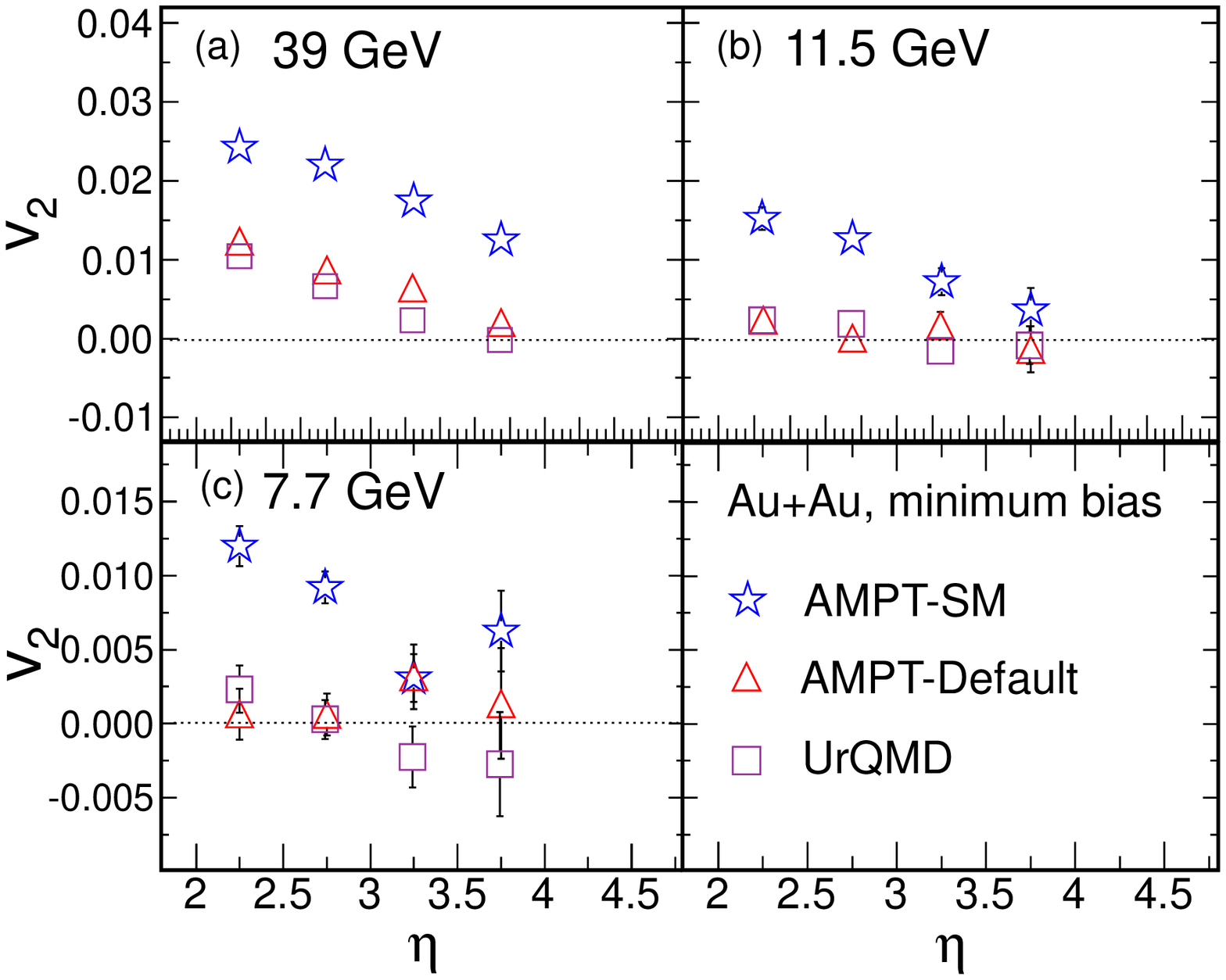}
\caption{(Color online) Expectation of inclusive photon $v_{2}$ as a function
collision centrality and $\eta$ (range where the measurements will be carried out in
actual experiment) for $\sqrt{s_{\mathrm {NN}}}$ = 7.7, 11.5 and 39 GeV Au+Au collisions
using AMPT-SM, AMPT default and UrQMD models.}
\label{bes}
\end{center}
\eef

Figure~\ref{bes} shows the inclusive photon $v_{2}$ expected from transport models at the
BES energies of 7.7, 11.5 and 39 GeV as a function of collision centrality and
pseudorapidity range where such measurements are possible in the experiment.
Not much difference is observed between the $v_{2}$ from AMPT default and UrQMD
at these energies, while the AMPT-SM results are significantly higher. These
results can be compared to data when available to understand the transition energy
from partonic to pure hadronic interaction as the dominant source of $v_{2}$~\cite{nasim2}.
The expected decrease in $v_{2}$ as the collision become more central or the $\eta$ increases
is observed for all the beam energies. Although at $\sqrt{s_{\mathrm {NN}}}$ = 7.7 and 11.5 GeV
AMPT-default and UrQMD predicts a negligible $v_{2}$ for the inclusive photon.

\bef
\begin{center}
\includegraphics[scale=0.4]{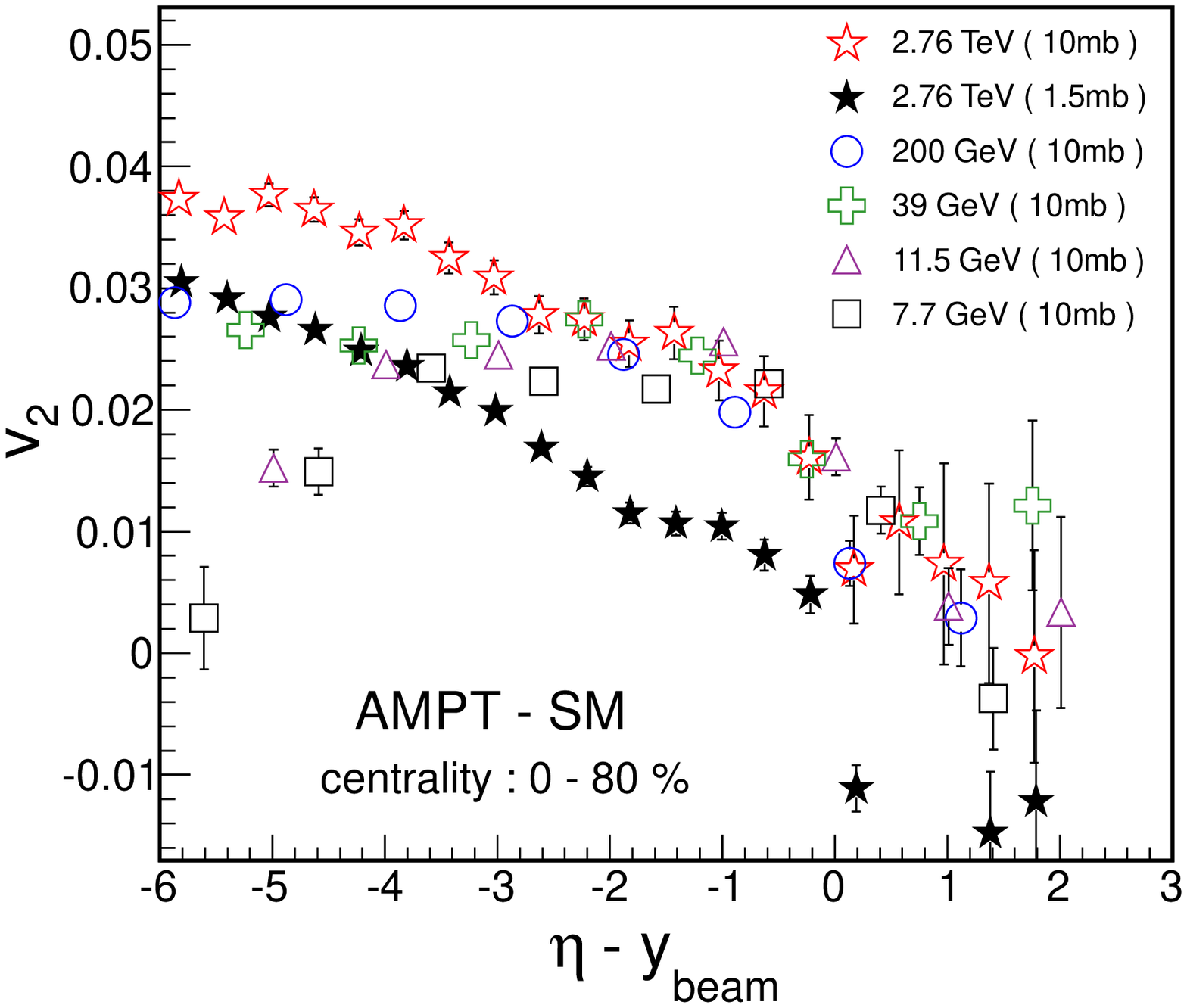}
\caption{(Color online) Inclusive photon $v_{2}$ as a function $\eta$ - $y_{\mathrm {beam}}$
for Au+Au collisions at $\sqrt{s_{\mathrm {NN}}}$ = 7.7, 11.5, 39 and 200 GeV with
parton cross section of 10 mb and Pb+Pb collisions at $\sqrt{s_{\mathrm {NN}}}$ = 2.76 TeV
for parton cross section of 1.5 and 10 mb. The results are shown for 0-80\% collision
centrality.}
\label{lf}
\end{center}
\eef
Figure~\ref{lf} shows the $v_{2}$ for inclusive photon as a
function $\eta$-$y_{\mathrm {beam}}$
for $\sqrt{s_{\mathrm {NN}}}$ = 7.7 GeV to 2.76 TeV using AMPT-SM with default settings.
The parton cross section is taken as 10 mb.
We observe the usual longitudinal scaling of $v_{2}$. We do not present
the results for the other models (AMPT-default and UrQMD) as those models have been shown
not to exhibit the longitudinal scaling behavior for $v_{2}$~\cite{nasim1}.
Recently it was shown that although the results with a partonic cross section
of 10 mb in AMPT-SM at RHIC is able to explain the measured distributions, for
those at LHC energies one requires a smaller cross section of 1.5 mb. In Fig.~\ref{lf}
we also show the $v_{2}$ vs. $\eta$-$y_{\mathrm {beam}}$
from AMPT-SM using cross section of 1.5mb at $\sqrt{s_{\mathrm {NN}}}$ = 2.76 TeV. We find
the longitudinal scaling behavior is violated. If the collisions at lower and
higher energies require different cross sections for explaining the measured distributions,
we would expect a breakdown of the longitudinal scaling. The photon $v_{2}$
measurements at RHIC BES and LHC are crucial for verifying this conclusion.

\section{Summary}

Keeping in mind the near future measurements of inclusive photon $v_{2}$
in RHIC BES and LHC programs, we have presented a transport model (UrQMD and AMPT)
based discussion on various expectations from such a measurement. We find that although
the $p_{\mathrm T}$ dependence of $v_{2}$ of inclusive photon in the transport models is
similar to those for the $\pi^{0}$ (90\% of the inclusive photon are from $\pi^{0}$ decay),
there is about 40\% decrease in the $p_{\mathrm T}$ integrated $v_{2}$ of photons relative
to those from $\pi^{0}$ for $\sqrt{s_{\mathrm {NN}}}$ = 7.7 GeV to 2.76 TeV. From a simple
modeling of the finite efficiency and purity of the counted photon sample, we find efficiency
has negligible effect on the measured $v_{2}$, however smaller purity tends to increase
the $v_{2}$ of inclusive photon. The later is because of the charged hadrons, which acts
as a source of contamination, are expected to have a higher $v_{2}$ than those from photons.

We find that the existing measurements of inclusive photon $v_{2}$ at RHIC energy
of $\sqrt{s_{\mathrm {NN}}}$ = 200 GeV is best explained using AMPT-SM, while
results from the AMPT default and UrQMD are smaller. This indicates that partonic interactions
are relevant for generation of $v_{2}$ even at rapidities away from midrapidity at top
RHIC energy. Such contributions appear to be much smaller at SPS energies, seen
from the comparison of $v_{2}$ data to AMPT default and UrQMD models. Comparison of the
centrality dependence of inclusive photon $v_{2}$ at SPS and RHIC energies to AMPT-SM model
calculations with different partonic cross sections of 3, 6 and 10 mb, indicates that the data
seems to favor a partonic cross section of 10 mb at both the energies.

We have provided the expected inclusive photon $v_{2}$ values at RHIC BES energies
using the transport models. The AMPT default and UrQMD models give smaller $v_{2}$ compared
to AMPT-SM, they also predict that the hadronic interactions at forward rapidities
may not be sufficient to generate substantial inclusive photon $v_{2}$ at
$\sqrt{s_{\mathrm {NN}}}$ = 7.7 GeV. The AMPT-SM model predicts that we should observe a
longitudinal scaling in inclusive photon $v_{2}$ from RHIC BES to LHC energies provided
the parton-parton cross section is independent of beam energy. A different parton-parton
cross section will lead to breaking of the longitudinal scaling in $v_{2}$.

\noindent{\bf Acknowledgments}\\
We thank Zi-Wei Lin for useful discussions on AMPT model results and
members of STAR collaboration for discussion on data presented.
Financial assistance from the Department of Atomic Energy, Government
of India is gratefully acknowledged. This work is supported by
the DAE-BRNS project sanction No. 2010/21/15-BRNS/2026.\\
%
%

\end{document}